# Local topological charge analysis of electromagnetic vortex beam based on empirical mode decomposition

Xiaonan Hui[1], Shilie Zheng,[1,a)] Weite Zhang,[1] Xiaofeng Jin,[1] Hao Chi,[1] and Xianmin Zhang[1,b)]

[1]*College of Information Science and Electronic Engineering, Zhejiang University, 310027, Hangzhou, China*

The topological charge of an electromagnetic vortex beam depends on its wavefront helicity. For mixed vortex beams composed of several different coaxial vortices, the topological charge spectrum can be obtained by Fourier transform. However, the vortex beam is generally divergent and imperfect. It makes it significant to investigate the local topological charges, especially in radio frequency regime. Fourier transform based methods are restrained by the uncertainty principle and cannot achieve high angular resolution and mode resolution simultaneously. In this letter, an analysis method for local topological charges of vortex beams is presented based on the empirical mode decomposition (EMD). From EMD, the intrinsic mode functions (IMFs) can be obtained to construct the bases of the electromagnetic wave, and each local topological charge can be respectively defined. With this method the local value achieves both high resolution of azimuth angle and topological charge, meanwhile the amplitudes of each OAM modes are presented as well. The simulation and experimental results confirm the validity of the EMD based method.

## I. INTRODUCTION

Since Allen *et al.* recognized in 1992 that the light beams with the transverse azimuthal dependence of $\exp(-jm\varphi)$ carry orbital angular momentum (OAM) [1], this fundamental physical property of electromagnetic waves is widely investigated [2,3]. The electromagnetic waves carrying OAM are applied in many fields, such as micro-machines driving [4], atoms trapping [5], stimulated emission depletion microscopy [6,7], and rotational Doppler shift [8]. In the most of applications, the beam with single OAM mode is employed, the topological charge of the OAM beam can be easily obtained. Furthermore, unlike the spin angular momentum (SAM) only has two orthogonal states, OAM can have unbounded eigenstates, which casts a bright prospect of the OAM mode multiplexing for high transmission capacity communication [9,10]. It is well known that OAM-carrying beams have the features of helicoidal wave-front and an amplitude null in the center of the beam. Due to the OAM beam

---

[a)] zhengsl@zju.edu.cn
[b)] zhangxm@zju.edu.cn

divergence, the 'dark zone' around the amplitude null will become larger with beam traveling. The divergence will also be more serious with the increase of the topological charge *m*. The divergence and imperfection of the OAM beams makes it significant to investigate the local topological charges, especially in radio frequency regime. In the last decade, various means to sort and detect OAM mode or superposition modes have been proposed with whole angular aperture receiving scheme [11-13] or partial angular aperture receiving scheme [14]. However, the Fourier transform based de-multiplexing is basically restrained by the uncertainty principle [15]. It cannot achieve high angular resolution and mode resolution simultaneously. The empirical mode decomposition (EMD) is one process of Hilbert-Huang transform (HHT). Unlike the Fourier transform which is a priori-defined basis transformation, HHT is a posteriori-defined bases method, which is not limited by the uncertainty principle [16] and has been widely applied in many research fields[17-19].

In this letter, we demonstrate an analysis method for local topological charges of vortex beams with superposition OAM modes based on EMD. The bases of the electromagnetic wave can be constructed by the intrinsic mode functions (IMFs) from the EMD, and thus each local topological charge can be respectively defined. Both the azimuth angle and local topological charge can be obtained with high resolution, and the OAM spectrum can be defined. The radiation of the multi-ring resonator OAM antenna with multiple OAM modes is numerically simulated and experimentally measured, and the local topological charge spectra are analyzed. The results confirm the validity of the EMD based method.

## II. THEORETICAL PRINCIPLE

For an *n*-OAM modes superposed beam, the co-polarization electric field along the azimuth angle can be presented as

$$E(\varphi) = \sum_{i=1}^{n} A_i \exp(-jm_i\varphi) \quad (1)$$

where $A_i$ is the amplitude of OAM mode *i*, and $m_i$ is the topological charge. Under the Fourier transform based OAM spectrum definition, the spectrum can be presented as

$$S(m_k) = \left| \int_{-\pi}^{\pi} E(\varphi) \exp[-j(-m_k)\varphi] d\varphi \right| \quad (2)$$

where $m_k \in (-\infty, \infty)$. However, the information of azimuth angle will be lost under this definition. In order to obtain the azimuth angle information, the common method is to insert an angular aperture or employ the windowed Fourier transform. When only one OAM mode is considered, that is to say $n=1$ in the Eq. (1), the local OAM mode spectrum can be presented as

$$S(m_k) = \left| \int_{\varphi_0 - \Delta\varphi/2}^{\varphi_0 + \Delta\varphi/2} E(\varphi) \exp[-j(-m_k)\varphi] d\varphi \right| = A_1 \Delta\varphi \, \text{sinc}[(m_1 - m_k)\Delta\varphi/2] \qquad (3)$$

The envelope of the spectrum is the sinc function, as shown in Fig. 1. The red bar chart is the OAM mode spectrum of $m_1=2$ with the partial angular receiving aperture $\Delta\varphi=10°$. It is clear that the spectrum expands greatly, so that the topological charge can hardly be identified. The blue bar chart is the spectrum of the superposition OAM field of $m_i=2, -6$, and 17, with the partial angular receiving aperture $\Delta\varphi=10°$. It does not show any information of the topological charges. So, the Fourier transform based OAM spectrum definition is restricted by the uncertainty principle and hardly to achieve high OAM mode resolution and angular resolution simultaneously.

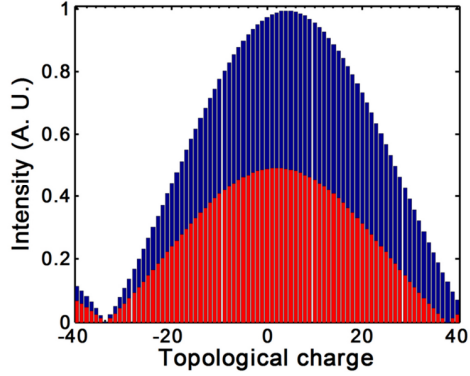

FIG. 1. The topological charge spectra with partial angular receiving aperture $\Delta\varphi=10°$. Red bar: the spectrum of $m=2$, and blue bar: the spectrum of superposed filed of $m=2, -6$, and 17.

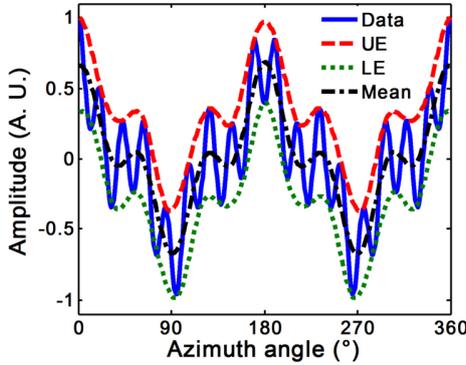

FIG. 2. An example of the real part of a superposition OAM field (blue solid line) to be analyzed. Upper envelop (UE, red dash line), lower envelop (LE, green dot line), and the mean value curve (black dash dot line).

Here, we demonstrate a new definition method. When $n=1$, the local topological charge $m(\varphi)$ and the corresponding amplitude $a(\varphi)$ can be easily defined as

$$m(\varphi) = \frac{d\theta(\varphi)}{d\varphi}, \text{ where } \theta(\varphi) = \arctan\left[\frac{E_{im}(\varphi)}{E_{re}(\varphi)}\right] \tag{4}$$

and

$$a(\varphi) = \sqrt{E_{re}(\varphi)^2 + E_{im}(\varphi)^2} \tag{5}$$

where $E_{re}(\varphi)$ and $E_{im}(\varphi)$ are the real part and the imaginary part of $E(\varphi)$, respectively. When $n>1$, there are multiple OAM modes superposition, the EMD process will be introduced to decompose $E_{re}(\varphi)$ and $E_{im}(\varphi)$. The real part of a superposition OAM field is shown as the blue solid curve in Fig. 2. The EMD can be operated by as follows. First, identify the local maximum values of $E_{re}(\varphi)$, and connect them with a cubic spline, rendered as the upper envelop (red dash line) in Fig. 2. Then, repeat this operation to the local minimum values. The lower envelope line (green dot line) can be obtained too. The mean value (black dash dot line) of the upper and lower envelopes is designated as $m_{11}$. The difference between $E_{re}(\varphi)$ and $m_{11}$ can be written as

$$h_{11} = E_{re}(\varphi) - m_{11} \tag{6}$$

In practice, $h_{11}$ can hardly fulfill the IMF definition: (a) the sum of maxima and minima in the data must either equal with the zero-crossings or differ at most by one; (b) the mean value of the two envelopes defined by the maxima and the minima is zero. So the sifting process should be operated. Treat $h_{11}$ as the proto-data, then identify the mean value of $h_{11}$ through the upper and lower envelops of $h_{11}$, and we can get $h_{12}$

$$h_{12} = h_{11} - m_{12} \tag{7}$$

where $m_{12}$ is the mean value obtained from $h_{11}$. After $k$ times, $h_{1k}$ meets the IMF definition (a) and (b), that is,

$$h_{1k} = h_{1(k-1)} - m_{1k} \tag{8}$$

Then, the first IMF of $E_{re}(\varphi)$ can be designated as $c_{re1}=h_{1k}$. Unlike the priori-defined basis such as the sine or cosine function with constant amplitude and vibration frequency, the IMF represents a simple oscillatory mode with variable amplitude and vibration frequency as a function of sample parameter ($\varphi$ in this case).

As $IMF_1$ is obtained, to continue the decomposition, and obtain the following IMFs, the original data subtracts the $IMF_1$, and $r_1$ is obtained.

$$r_{re1} = E_{re}(\varphi) - c_{re1} \tag{9}$$

Treat $r_{re1}$ as the data to be analyzed, and calculate the upper and lower envelops with the local maximum (minimum) values and the cubic spline lines. The mean value of $r_{re1}$ can be got, as $m_{21}$. Repeat the sifting processing described by Eq. (6) to Eq. (9), until the final residue $r_{ren}$ is either the mean trend or a constant.

So, all the IMFs ($c_{rei}$) can be obtained with this method, and the $E_{re}(\varphi)$ can be presented as

$$E_{re}(\varphi) = \sum_{i=1}^{n} c_{rei} + r_{ren} \tag{10}$$

Repeat the same process with the imaginary part $E_{im}(\varphi)$, and it can be presented as

$$E_{im}(\varphi) = \sum_{i=1}^{n} c_{imi} + r_{imn} \tag{11}$$

Combined with Eq. (4) and Eq. (5), for each pair of $c_{rei}$ and $c_{imi}$, the local mode and amplitude can be obtained as

$$m_i(\varphi) = \frac{d\theta_i(\varphi)}{d\varphi}, \text{ where } \theta_i(\varphi) = \arctan\left(\frac{c_{imi}}{c_{rei}}\right) \tag{12}$$

and

$$a_i(\varphi) = \sqrt{c_{rei}^2 + c_{imi}^2} \tag{13}$$

The field can be expressed as

$$E(\varphi) = \sum_{i=1}^{n} a_i(\varphi) \exp\left[-j \int m_i(\varphi) d\varphi\right] \tag{14}$$

here the residuals are not considered. After the process of EMD, the local topological charges of the superposed OAM beam can be defined and calculated by Eq. (12) to Eq. (14).

## III. SIMULATION AND EXPERIMENT

To verify the definition method, we use a superposition field, which contains the OAM modes of −6, 2, and 17 with the amplitude ratio of 2.5:4.5:3. The real part of the field is shown as the blue solid line in Fig. 2. The field is discretized by 360 samples with a step of 1°. The real part and the imaginary part are processed by the EMD respectively, and the results of IMFs are shown in Fig. 3. It can be seen that the data are mainly decomposed into 3 IMFs, namely 3 topological oscillations. The components with topological charges $m$=17, −6 and 2 are decomposed into IMF1, IMF2, and IMF3, respectively.

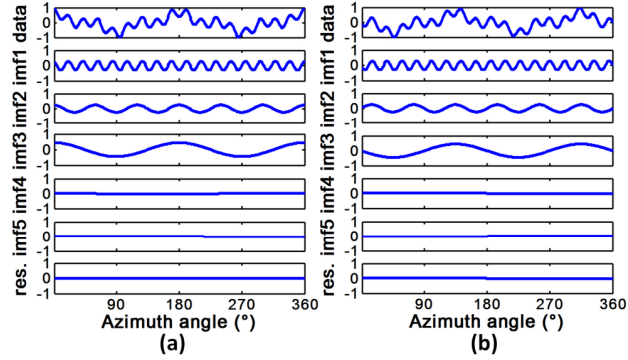

FIG. 3. The IMFs of the field. (a) The IMFs of the real part, and (b) the IMFs of the imaginary part.

With the definition of local topological charge, the analytic expression of the field can be written as Eq. (14), so the local topological charge spectrum can be calculated, as shown in Fig. 4(a). The local topological charge spectrum demonstrates the relationship of azimuth angle (horizon axis), calculated topological charge number (vertical axis), and the amplitude (color). It can be seen from Fig. 4(a), there are 3 horizontal lines representing the components of $m$=−6, 2, and 17 along the azimuth angle. The amplitudes denoted by the color show good agreement with the set ratio of 2.5:4.5:3. Therefore, the local topological charges can be identified from the spectrum easily, and the azimuth angle resolution depends on the sampling resolution (1° in this case). However, the local topological charges can be integrated along azimuth angle. Fig. 4(b) shows the integrated value within 10° from 171° to 180°. Compared with the result based on Fourier transform shown in Fig.1, the local topological charge spectrum presents a significant meaning for OAM mode analysis. It should be noted that the intensity peaks in Fig. 4(b) do not agree with the amplitudes' ratio of 2.5:4.5:3. The reason lies that the fluctuations of the calculated mode spectra (the yellow, red and cyan spectra in Fig. 4(a)) widen the mode peaks in different way, which makes the intensity of the mode peak different with the original amplitude ratio. However the integral values of each mode peaks agree well with that ratio.

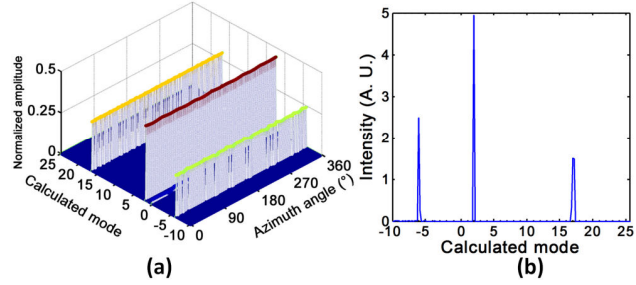

FIG. 4. The theoretical calculation of (a) the local topological charge spectrum, and (b) the 10° integrated local topological charge spectrum.

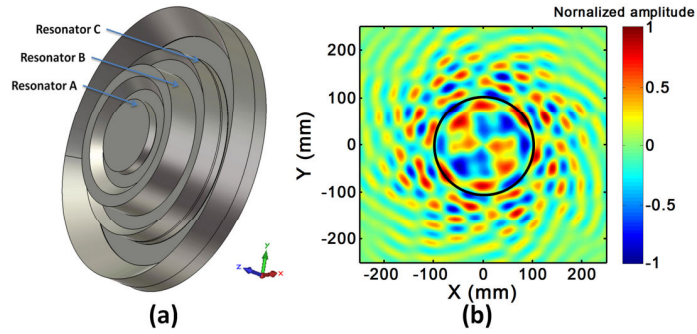

FIG. 5. The simulation model and the simulation field results. (a) The multi-resonator OAM antenna model, and (b) the simulation results of the real part of electric field in radial polarization.

To verify the calculation results, the simulation and experiment are performed. A multi-ring resonator OAM antenna with center frequency of 10 GHz is designed based on our previous works [20,21], as shown in Fig. 5(a). There are three resonators, namely resonators A, B, and C, respectively, which radiate the microwave beams carrying OAM modes $m$=2, −6, and 17, respectively. The radial polarization is assigned as the co-polarization. Fig. 5(b) is the simulation results by CST Microwave Studio. It shows the real part of electric field in radial polarization. The observation window is 500 mm × 500 mm lies in the plane 300 mm away from the antenna. The data are extracted along a circle with radius of 120 mm, shown as the black circle in Fig. 5(b). The amplitude ratio for these 3 OAM modes is also designed to be 2.5:4.5:3. After the calculation, the local topological charge spectrum is obtained and shown in Fig. 6(a). The spectrum exhibits the distribution of topological charge along the azimuth angle clearly, which fits well with the theoretically calculated spectrum shown in Fig. 4(a). Fig. 6(b) shows the 10° integrated local topological charge spectrum, which is also agreement with the theoretically calculated one shown in Fig. 4(b).

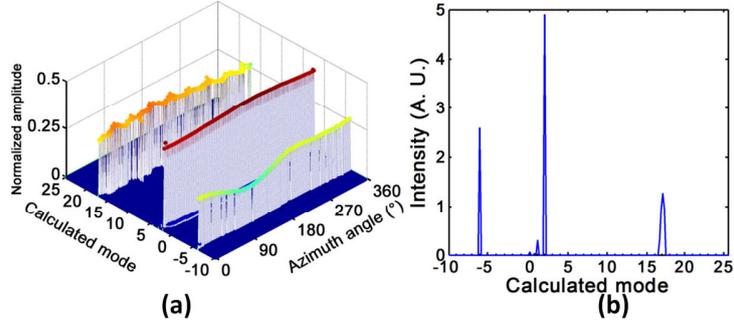

FIG. 6. The simulation results. (a) The local topological charge spectrum, and (b) the 10° integrated local topological charge spectrum.

Furthermore, an experiment is carried out. A two-resonator OAM antenna is designed and fabricated with copper, as shown in Fig. 7(a). A beam carrying OAM modes $m=-2$ and 3 can be generated. The resonator OAM antenna is settled on a rotation platform and the electric field along a circle is measured. The measurement plane is 400 mm away from the OAM antenna, and the circle radius is 200 mm. The radial polarization field is detected by an open-end waveguide, and the real part and the imaginary part of the fields are recorded by a vector network analyzer (R&S ZVA-67). The field is sampled with a step of 2°. Then, the data are processed with the method described above. The local topological charge spectrum is shown in Fig. 7(b). It is seen that the topological charges are mainly distributed at the modes $m=-2$ and 3. Because the field is sampled every 2°, the azimuth angular resolution achieves about 2°. The resolution of the calculated topological charge number depends on the calculation setting, which is 0.2 in this spectral diagram.

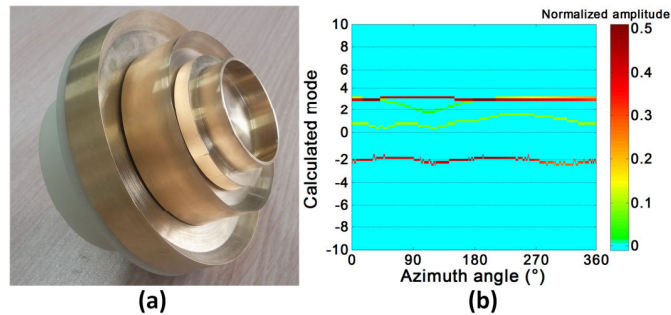

FIG. 7. The OAM antenna and experimental results. (a) Two-resonator OAM antenna, and (b) the local topological charge spectrum.

## IV. CONCLUSION

It is significant to investigate the local topological charges of OAM beam due to its divergence and imperfection, especially in radio frequency regime. We demonstrate a method to define and calculate the local topological charges of the superposed multi-mode OAM beam. The EMD method decomposes the field into the IMFs which represent the variation of the field along the azimuth angle. Based on the EMD method, the local topological charge spectrum can be calculated. In this letter, the angular resolution achieves as high as 1° in the simulation and 2° in the experiment, and the topological charge number resolution is set as 0.2. Compared with the Fourier transform, the EMD based method is not limited by the uncertainty principle and can achieve high angular and mode resolution simultaneously. It offers a powerful analysis tool for future investigation and applications of electromagnetic orbital angular momentum.

## ACKNOWLEDGMENTS

This work was supported by the National Basic Research Program of China (973 program) under Grant 2014CB340005 and 2014CB340001, and the Natural Science Foundation of China under Grant 61371030.